\documentclass[twocolumn,letterpaper,aps,prl,longbibliography,superscriptaddress,floatfix]{revtex4-2}

\usepackage{graphicx}   
\usepackage{ltxtable}
\usepackage{amsmath}
\usepackage{verbatim}
\usepackage{xspace}     
\usepackage{float}

\usepackage{xspace}     
\usepackage{enumitem}
\setlist[enumerate]{align=left}

\newcommand{\pT}{\mbox{$p_T$}\xspace}
\newcommand{\pizero}{\mbox{$\pi^0$}\xspace}
\newcommand{\ALL}{\mbox{$A_{LL}$}\xspace}

\begin{document}

\title{Measurement of Direct-Photon Cross Section and Double-Helicity 
Asymmetry at $\sqrt{s}=510$ GeV in $\vec{p}+\vec{p}$ Collisions}

\newcommand{\abilene}{Abilene Christian University, Abilene, Texas 79699, USA}
\newcommand{\augie}{Department of Physics, Augustana University, Sioux Falls, South Dakota 57197, USA}
\newcommand{\banaras}{Department of Physics, Banaras Hindu University, Varanasi 221005, India}
\newcommand{\barc}{Bhabha Atomic Research Centre, Bombay 400 085, India}
\newcommand{\baruch}{Baruch College, City University of New York, New York, New York, 10010 USA}
\newcommand{\bnlcoll}{Collider-Accelerator Department, Brookhaven National Laboratory, Upton, New York 11973-5000, USA}
\newcommand{\bnlphys}{Physics Department, Brookhaven National Laboratory, Upton, New York 11973-5000, USA}
\newcommand{\caucr}{University of California-Riverside, Riverside, California 92521, USA}
\newcommand{\charlesczech}{Charles University, Faculty of Mathematics and Physics, 180 00 Troja, Prague, Czech Republic}
\newcommand{\cns}{Center for Nuclear Study, Graduate School of Science, University of Tokyo, 7-3-1 Hongo, Bunkyo, Tokyo 113-0033, Japan}
\newcommand{\colorado}{University of Colorado, Boulder, Colorado 80309, USA}
\newcommand{\columbia}{Columbia University, New York, New York 10027 and Nevis Laboratories, Irvington, New York 10533, USA}
\newcommand{\czechtech}{Czech Technical University, Zikova 4, 166 36 Prague 6, Czech Republic}
\newcommand{\debrecen}{Debrecen University, H-4010 Debrecen, Egyetem t{\'e}r 1, Hungary}
\newcommand{\elte}{ELTE, E{\"o}tv{\"o}s Lor{\'a}nd University, H-1117 Budapest, P{\'a}zm{\'a}ny P.~s.~1/A, Hungary}
\newcommand{\ewha}{Ewha Womans University, Seoul 120-750, Korea}
\newcommand{\famu}{Florida A\&M University, Tallahassee, FL 32307, USA}
\newcommand{\fsu}{Florida State University, Tallahassee, Florida 32306, USA}
\newcommand{\gsu}{Georgia State University, Atlanta, Georgia 30303, USA}
\newcommand{\hanyang}{Hanyang University, Seoul 133-792, Korea}
\newcommand{\hiroshima}{Hiroshima University, Kagamiyama, Higashi-Hiroshima 739-8526, Japan}
\newcommand{\howard}{Department of Physics and Astronomy, Howard University, Washington, DC 20059, USA}
\newcommand{\ihepprot}{IHEP Protvino, State Research Center of Russian Federation, Institute for High Energy Physics, Protvino, 142281, Russia}
\newcommand{\illuiuc}{University of Illinois at Urbana-Champaign, Urbana, Illinois 61801, USA}
\newcommand{\inrras}{Institute for Nuclear Research of the Russian Academy of Sciences, prospekt 60-letiya Oktyabrya 7a, Moscow 117312, Russia}
\newcommand{\instpasczech}{Institute of Physics, Academy of Sciences of the Czech Republic, Na Slovance 2, 182 21 Prague 8, Czech Republic}
\newcommand{\isu}{Iowa State University, Ames, Iowa 50011, USA}
\newcommand{\jaea}{Advanced Science Research Center, Japan Atomic Energy Agency, 2-4 Shirakata Shirane, Tokai-mura, Naka-gun, Ibaraki-ken 319-1195, Japan}
\newcommand{\jeonbuk}{Jeonbuk National University, Jeonju, 54896, Korea}
\newcommand{\jyvaskyla}{Helsinki Institute of Physics and University of Jyv{\"a}skyl{\"a}, P.O.Box 35, FI-40014 Jyv{\"a}skyl{\"a}, Finland}
\newcommand{\kek}{KEK, High Energy Accelerator Research Organization, Tsukuba, Ibaraki 305-0801, Japan}
\newcommand{\korea}{Korea University, Seoul 02841, Korea}
\newcommand{\kurchatov}{National Research Center ``Kurchatov Institute", Moscow, 123098 Russia}
\newcommand{\kyoto}{Kyoto University, Kyoto 606-8502, Japan}
\newcommand{\lahorelums}{Physics Department, Lahore University of Management Sciences, Lahore 54792, Pakistan}
\newcommand{\lawllnl}{Lawrence Livermore National Laboratory, Livermore, California 94550, USA}
\newcommand{\losalamos}{Los Alamos National Laboratory, Los Alamos, New Mexico 87545, USA}
\newcommand{\lund}{Department of Physics, Lund University, Box 118, SE-221 00 Lund, Sweden}
\newcommand{\maryland}{University of Maryland, College Park, Maryland 20742, USA}
\newcommand{\mass}{Department of Physics, University of Massachusetts, Amherst, Massachusetts 01003-9337, USA}
\newcommand{\mate}{MATE, Laboratory of Femtoscopy, K\'aroly R\'obert Campus, H-3200 Gy\"ongy\"os, M\'atrai\'ut 36, Hungary}
\newcommand{\michigan}{Department of Physics, University of Michigan, Ann Arbor, Michigan 48109-1040, USA}
\newcommand{\miss}{Mississippi State University, Mississippi State, Mississippi 39762, USA}
\newcommand{\muhlenberg}{Muhlenberg College, Allentown, Pennsylvania 18104-5586, USA}
\newcommand{\myongji}{Myongji University, Yongin, Kyonggido 449-728, Korea}
\newcommand{\nara}{Nara Women's University, Kita-uoya Nishi-machi Nara 630-8506, Japan}
\newcommand{\natmephi}{National Research Nuclear University, MEPhI, Moscow Engineering Physics Institute, Moscow, 115409, Russia}
\newcommand{\newmex}{University of New Mexico, Albuquerque, New Mexico 87131, USA}
\newcommand{\nmsu}{New Mexico State University, Las Cruces, New Mexico 88003, USA}
\newcommand{\northcg}{Physics and Astronomy Department, University of North Carolina at Greensboro, Greensboro, North Carolina 27412, USA}
\newcommand{\ohio}{Department of Physics and Astronomy, Ohio University, Athens, Ohio 45701, USA}
\newcommand{\ornl}{Oak Ridge National Laboratory, Oak Ridge, Tennessee 37831, USA}
\newcommand{\orsay}{IPN-Orsay, Univ.~Paris-Sud, CNRS/IN2P3, Universit\'e Paris-Saclay, BP1, F-91406, Orsay, France}
\newcommand{\pnpi}{PNPI, Petersburg Nuclear Physics Institute, Gatchina, Leningrad region, 188300, Russia}
\newcommand{\pusan}{Pusan National University, Pusan 46241, Korea}
\newcommand{\riken}{RIKEN Nishina Center for Accelerator-Based Science, Wako, Saitama 351-0198, Japan}
\newcommand{\rikjrbrc}{RIKEN BNL Research Center, Brookhaven National Laboratory, Upton, New York 11973-5000, USA}
\newcommand{\rikkyo}{Physics Department, Rikkyo University, 3-34-1 Nishi-Ikebukuro, Toshima, Tokyo 171-8501, Japan}
\newcommand{\saispbstu}{Saint Petersburg State Polytechnic University, St.~Petersburg, 195251 Russia}
\newcommand{\seoulnat}{Department of Physics and Astronomy, Seoul National University, Seoul 151-742, Korea}
\newcommand{\stonybrkc}{Chemistry Department, Stony Brook University, SUNY, Stony Brook, New York 11794-3400, USA}
\newcommand{\stonycrkp}{Department of Physics and Astronomy, Stony Brook University, SUNY, Stony Brook, New York 11794-3800, USA}
\newcommand{\tenn}{University of Tennessee, Knoxville, Tennessee 37996, USA}
\newcommand{\texsu}{Texas Southern University, Houston, TX 77004, USA}
\newcommand{\titech}{Department of Physics, Tokyo Institute of Technology, Oh-okayama, Meguro, Tokyo 152-8551, Japan}
\newcommand{\tsukuba}{Tomonaga Center for the History of the Universe, University of Tsukuba, Tsukuba, Ibaraki 305, Japan}
\newcommand{\vandy}{Vanderbilt University, Nashville, Tennessee 37235, USA}
\newcommand{\weizmann}{Weizmann Institute, Rehovot 76100, Israel}
\newcommand{\wigner}{Institute for Particle and Nuclear Physics, Wigner Research Centre for Physics, Hungarian Academy of Sciences (Wigner RCP, RMKI) H-1525 Budapest 114, POBox 49, Budapest, Hungary}
\newcommand{\yonsei}{Yonsei University, IPAP, Seoul 120-749, Korea}
\newcommand{\zagreb}{Department of Physics, Faculty of Science, University of Zagreb, Bijeni\v{c}ka c.~32 HR-10002 Zagreb, Croatia}
\newcommand{\zambia}{Department of Physics, School of Natural Sciences, University of Zambia, Great East Road Campus, Box 32379, Lusaka, Zambia}
\affiliation{\abilene}
\affiliation{\augie}
\affiliation{\banaras}
\affiliation{\barc}
\affiliation{\baruch}
\affiliation{\bnlcoll}
\affiliation{\bnlphys}
\affiliation{\caucr}
\affiliation{\charlesczech}
\affiliation{\cns}
\affiliation{\colorado}
\affiliation{\columbia}
\affiliation{\czechtech}
\affiliation{\debrecen}
\affiliation{\elte}
\affiliation{\ewha}
\affiliation{\famu}
\affiliation{\fsu}
\affiliation{\gsu}
\affiliation{\hanyang}
\affiliation{\hiroshima}
\affiliation{\howard}
\affiliation{\ihepprot}
\affiliation{\illuiuc}
\affiliation{\inrras}
\affiliation{\instpasczech}
\affiliation{\isu}
\affiliation{\jaea}
\affiliation{\jeonbuk}
\affiliation{\jyvaskyla}
\affiliation{\kek}
\affiliation{\korea}
\affiliation{\kurchatov}
\affiliation{\kyoto}
\affiliation{\lahorelums}
\affiliation{\lawllnl}
\affiliation{\losalamos}
\affiliation{\lund}
\affiliation{\maryland}
\affiliation{\mass}
\affiliation{\mate}
\affiliation{\michigan}
\affiliation{\miss}
\affiliation{\muhlenberg}
\affiliation{\myongji}
\affiliation{\nara}
\affiliation{\natmephi}
\affiliation{\newmex}
\affiliation{\nmsu}
\affiliation{\northcg}
\affiliation{\ohio}
\affiliation{\ornl}
\affiliation{\orsay}
\affiliation{\pnpi}
\affiliation{\pusan}
\affiliation{\riken}
\affiliation{\rikjrbrc}
\affiliation{\rikkyo}
\affiliation{\saispbstu}
\affiliation{\seoulnat}
\affiliation{\stonybrkc}
\affiliation{\stonycrkp}
\affiliation{\tenn}
\affiliation{\texsu}
\affiliation{\titech}
\affiliation{\tsukuba}
\affiliation{\vandy}
\affiliation{\weizmann}
\affiliation{\wigner}
\affiliation{\yonsei}
\affiliation{\zagreb}
\affiliation{\zambia}
\author{N.J.~Abdulameer} \affiliation{\debrecen}
\author{U.~Acharya} \affiliation{\gsu} 
\author{A.~Adare} \affiliation{\colorado} 
\author{C.~Aidala} \affiliation{\michigan} 
\author{N.N.~Ajitanand} \altaffiliation{Deceased} \affiliation{\stonybrkc} 
\author{Y.~Akiba} \email[PHENIX Spokesperson: ]{akiba@rcf.rhic.bnl.gov} \affiliation{\riken} \affiliation{\rikjrbrc} 
\author{R.~Akimoto} \affiliation{\cns} 
\author{M.~Alfred} \affiliation{\howard} 
\author{N.~Apadula} \affiliation{\isu} \affiliation{\stonycrkp} 
\author{Y.~Aramaki} \affiliation{\riken} 
% ...
\collaboration{PHENIX Collaboration}  \noaffiliation

\date{\today}


\begin{abstract}


We present measurements of the cross section and double-helicity 
asymmetry $A_{LL}$ of direct-photon production in $\vec{p}+\vec{p}$ 
collisions at $\sqrt{s}=510$ GeV. The measurements have been performed 
at midrapidity ($|\eta|<0.25$) with the PHENIX detector at the 
Relativistic Heavy Ion Collider.  At relativistic energies, direct 
photons are dominantly produced from the initial quark-gluon hard 
scattering and do not interact via the strong force at leading order. 
Therefore, at $\sqrt{s}=510$ GeV, where leading-order-effects dominate, 
these measurements provide clean and direct access to the gluon helicity 
in the polarized proton in the gluon-momentum-fraction range 
$0.02<x<0.08$, with direct sensitivity to the sign of the gluon 
contribution.

\end{abstract}

\maketitle




In polarized-proton collisions, spin-asymmetry measurements are 
sensitive to the polarized partonic structure of the proton and allow 
the investigation of its spin decomposition. Determining how fundamental 
properties of a particle such as spin comprise its constituents 
is of great importance in understanding quantum chromodynamics (QCD). 
Perturbative QCD (pQCD) has been successful in describing unpolarized 
cross sections while spin-dependent observables have historically 
offered additional insights. Polarized deep-inelastic scattering (DIS) 
has shown that only part of the proton spin is carried by quarks.
A large fraction of the proton spin was suggested to be carried by 
gluons~\cite{1988364,ASHMAN19891,PhysRevD.58.112001,ALEXAKHIN20078, 
PhysRevD.75.012007}. DIS is sensitive to gluons only through high-order 
interactions and the polarized gluon distribution is significantly less 
constrained compared to the unpolarized gluon due to the (so far) limited 
kinematic coverage of polarized data. At the Relativistic Heavy 
Ion Collider (RHIC), gluons are accessible at leading order in the hard 
scattering.  Measurements of the double-helicity asymmetry (\ALL) are 
directly sensitive to the polarized gluon distribution via 
longitudinally polarized $\vec{p}+\vec{p}$ collisions. Recent RHIC 
measurements of \pizero and jets at $\sqrt{s}$ = 62.4 and 200 
GeV~\cite{PhysRevD.90.012007,PhysRevLett.103.012003,PhysRevD.79.012003, 
PhysRevD.86.032006,PhysRevLett.115.092002} that 
were included in global analyses have shown the first direct evidence of 
nonzero gluon-spin contributions to the spin of the 
proton~\cite{PhysRevLett.113.012001, 2014276} in the gluon momentum 
fraction ($x$) range larger than 0.05. Measurements at the higher energy 
of $\sqrt{s}$ = 510 GeV~\cite{PhysRevD.93.011501,PhysRevD.100.052005} 
have confirmed the nonzero gluon polarization and extended the minimum 
$x$ reach to $\approx$0.01. Recent analysis by the Jefferson Lab Angular 
Momentum (JAM) Collaboration showed that the two scenarios of positive 
and negative gluon-spin contributions are indistinguishable from each 
other based on the existing data~\cite{Zhou:2022wzm,PhysRevD.106.L031502}.
This can be resolved using direct-photon production in 
$\vec{p}+\vec{p}$ scattering, which is linearly sensitive to gluon 
helicity.

Direct photons are all those photons that are not coming from decays of 
final-state hadrons. The quark-gluon Compton process $qg \rightarrow 
q\gamma$ in proton-proton collisions at RHIC is the dominant contributor 
to the direct photons with transverse momentum larger than 5 GeV/$c$. 
Unlike hadrons and jets, direct photons do not involve color 
interactions in the final state. Therefore, they provide a direct probe 
to the initial state of colliding protons. The double-helicity asymmetry 
of direct-photon production in longitudinally polarized $\vec{p}+\vec{p}$ 
collisions is sensitive to both the sign and magnitude of the gluon-spin 
contributions to the proton spin. For this reason, \ALL was thought 
to be a \textit{golden} channel to access the gluon spin in the 1992 
RHIC-spin proposal~\cite{Bunce:1992vca,doi:10.1146/annurev.nucl.50.1.525}. 
In this Letter, we report the first measurements of this observable.


The data were collected in 2013 with the PHENIX 
detector at RHIC~\cite{ADCOX2003469} at $\sqrt{s}=510$ GeV within 
pseudorapidity $|\eta|<0.25$. We have extracted the inclusive and 
isolated direct-photon cross sections and \ALL of isolated photons. The 
primary detector for this measurement is an electromagnetic calorimeter 
(EMCal)~\cite{APHECETCHE2003521} comprising two subsystems, a 
six-sector lead-scintillator (PbSc) detector, of which four are on the 
west arm and two on the east arm, and a two-sector lead glass (PbGl) 
detector on the east arm, each located 5 m radially from the beam line. 
Each sector covers a range of $|\eta| <$ 0.35 and $22.5^{\rm o}$ in 
azimuth $\phi$. The EMCal has fine granularity with each tower covering 
$\Delta\eta \times \Delta\phi \approx$ 0.011 $\times$ 0.011 (0.008 
$\times$ 0.008) for PbSc (PbGl). Two photons from $\pi^0 \rightarrow 
\gamma\gamma$ decays are fully resolved up to a \pizero \pT of 12 (16) 
GeV/$c$ in the PbSc (PbGl), and a shower profile analysis extends the 
$\gamma/\pi^0$ discrimination up to 30 GeV/$c$ in these measurements. 
The energy calibration of each tower is obtained from the reconstructed 
\pizero mass.

The beam-beam counters (BBC)~\cite{ALLEN2003549} cover \mbox{3.1 $<
|\eta| <$ 3.9} and are located at $\pm$144 cm from the interaction point 
along the beam line. The BBCs measure the longitudinal collision vertex 
and provide a minimum-bias trigger. The BBCs are also used as 
a luminosity ($\mathcal{L}$) monitor. Events with high-\pT photons are selected by an 
EMCal-based trigger requiring a minimum energy deposit of 3.7 GeV in an 
overlapping tile of $4{\times}4$ towers of the EMCal in coincidence with 
the minimum-bias trigger. The cross-section (\ALL) analysis uses an 
integrated luminosity of 11 (108) pb$^{-1}$ with a 
z-vertex requirement of 10 (30) cm around the nominal interaction point. 
The photon-reconstruction and analysis method used here is 
similar to the previous PHENIX measurement at 
$\sqrt{s}=200$~GeV~\cite{PhysRevLett.98.012002,PhysRevD.86.072008}. 
Photons are identified by a shower-profile requirement that was 
calibrated using test-beam data, identified electrons, and decay photons 
from identified \pizero. The method rejects $\approx$50\% of hadrons 
depositing \mbox{$E>$ 3 GeV} in the EMCal and accepts $\approx$98\% of 
real photons. The time-of-flight (ToF) of particles is measured relative 
to the photon signal in the EMCal.  A ToF requirement 
$|{\rm ToF}|<10$~ns is used to reduce pileup events due to high 
collision rate (the average number of BBC triggered events per beam 
crossing varied in the range 0.04--0.17). A minimum-energy requirement 
$E_{\rm min}>0.3$ GeV is applied for the EMCal clusters to reduce the 
background noise.  The charged-particle veto of the photon sample is 
based on tracks in the drift chambers~\cite{ADCOX2003489}.


The experimental challenge in this measurement is the large photon 
background from hadron decays, primarily from 
$\pi^0 \rightarrow \gamma\gamma$ ($\approx$80\% of the decays) 
and $\eta \rightarrow 
\gamma\gamma$ ($\approx$15\%). Photon candidates that form a pair with 
another photon in the mass range 
$110<M_{\gamma\gamma}<160$~MeV/c$^2$ ($M_{\pi^0}{\pm}3\sigma$) 
with $E_{\gamma}>300$~MeV are 
tagged as \pizero decay photons. A fiducial region for direct-photon 
candidates excludes 10 (12) towers (0.1 rad) from the edges of the PbSc 
(PbGl). Partner photons are accepted over the entire detector to 
improve the probability of observing both decay photons from the 
\pizero.  This method overestimates $\approx$8\% more yield of photons 
from \pizero decays, $\gamma_{\pi^0}^{{\rm inc}}$, due to combinatorial 
background. A \pT-dependent correction is estimated from the fit of the 
background under the \pizero peak in the two-photon invariant-mass 
distribution. The inclusive direct-photon yield is then determined as

\begin{equation} \label{eq:inc}
\gamma_{{\rm dir}}^{{\rm inc}} = \gamma_{{\rm total}}^{{\rm inc}} - \left( 1 + R_{\pi^0}^{{\rm miss}} + \delta_{h/\pi^0}^{\gamma} \right) \gamma_{\pi^0}^{{\rm inc}},
\end{equation}

\noindent where we subtract the reconstructed inclusive photons from \pizero decay 
($\gamma_{\pi^0}^{{\rm inc}}$), those missing their partner photons 
($R_{\pi^0}^{{\rm miss}}\gamma_{\pi^0}^{{\rm inc}}$) and photons from 
other hadron decays 
($\delta_{h/\pi^0}^{\gamma}\gamma_{\pi^0}^{{\rm inc}}$) from the total 
inclusive photon sample ($\gamma_{{\rm total}}^{{\rm inc}}$). If a 
partner photon of a \pizero decay is missed, it will not be 
reconstructed in the \pizero mass peak window. The ratio of \pizero 
decay photons that missed their partner photons to those that were 
reconstructed, $R_{\pi^0}^{{\rm miss}}$, is estimated using a single 
\pizero simulation with photon shower and detector geometry. 
The $\delta_{h/\pi^0}^{\gamma}$ is calculated by $\eta$, $\omega$, $\eta'$ 
over \pizero ratios based on the previous $\sqrt{s}$ = 200 GeV 
measurement~\cite{PhysRevD.83.052004}: 
$\delta_{h/\pi^0}^{\gamma}{\approx}0.28$, with 
$\delta_{\eta/\pi^0}^{\gamma}{\approx}0.21$ and 
$\delta_{\omega/\pi^0}^{\gamma}{\approx}\delta_{\eta'/\pi^0}^{\gamma}{\approx}0.035$. 
A {\sc pythia}~\cite{Sjostrand:2006za} simulation showed that 
the variation of these ratios is less than 10\% between 200 GeV and 510 
GeV within 6 $< p_T <$ 30 GeV/$c$. The difference is accounted for by 
assigning a systematic uncertainty.

\begin{figure*}[hbt!]
\includegraphics[width=0.48\linewidth]{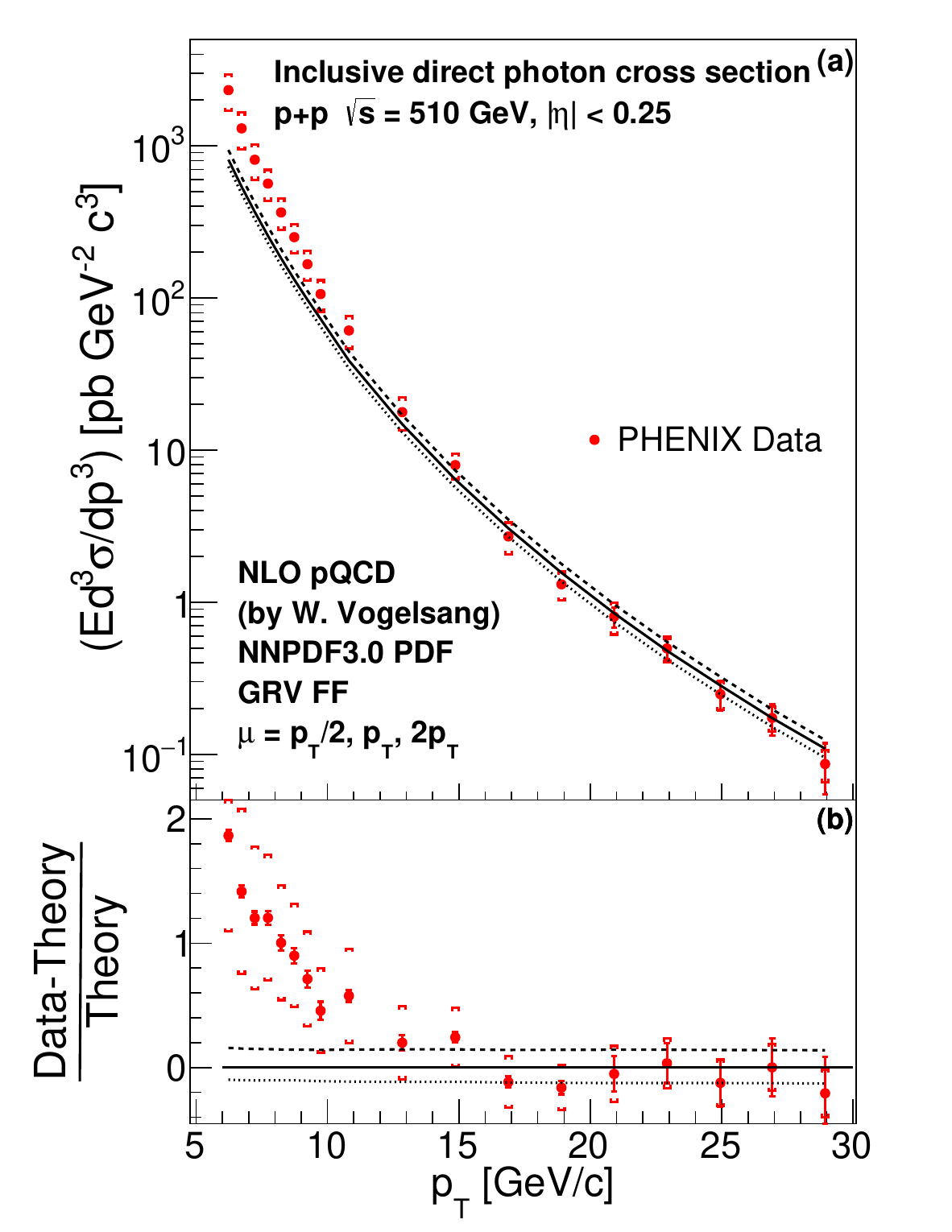}
\includegraphics[width=0.48\linewidth]{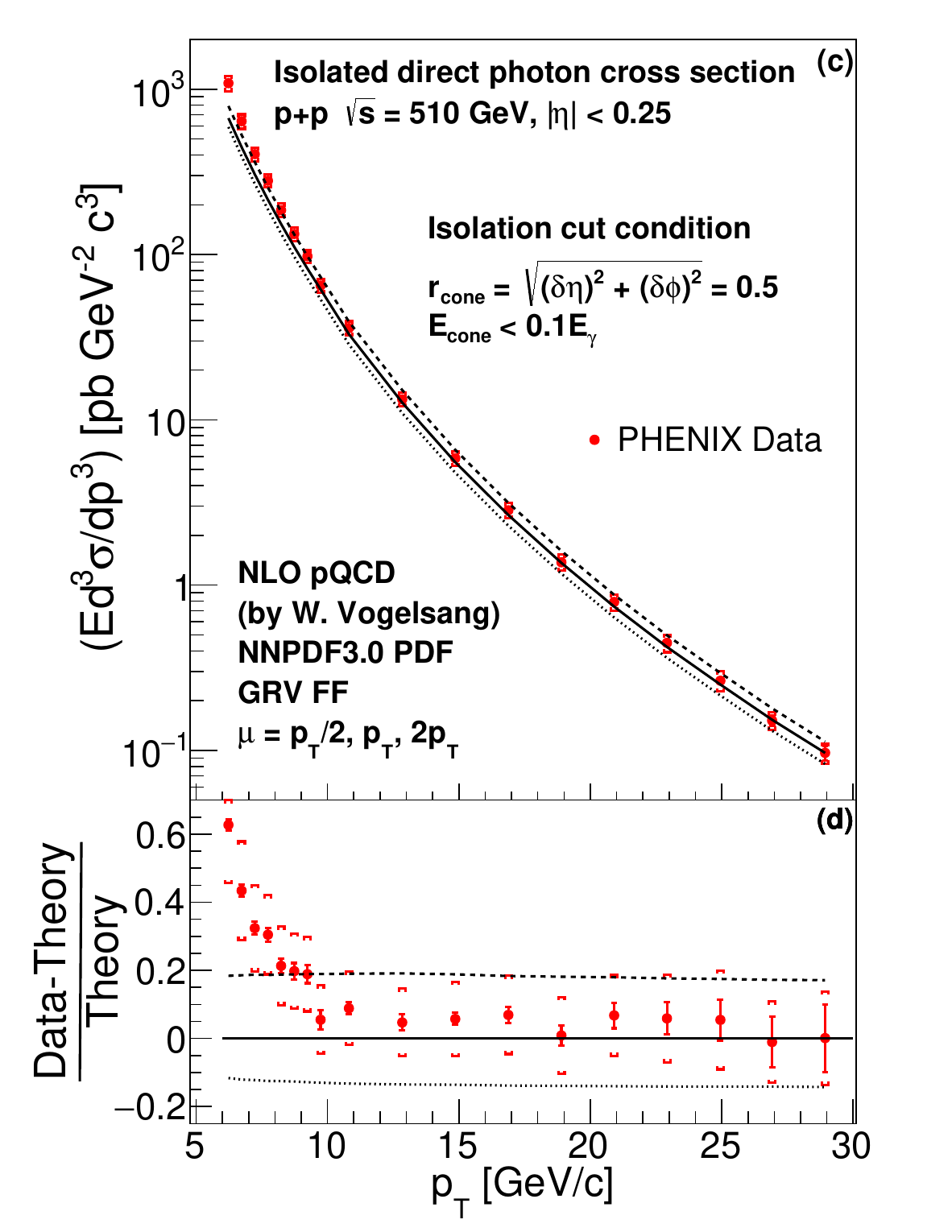}
\caption{Cross sections for (a) inclusive and (c) isolated direct 
photons as a function of \pT compared with next-to-leading-order (NLO) 
pQCD calculations~\protect\cite{PhysRevD.48.3136,PhysRevD.50.1901} for 
different renormalization and factorization scales $\mu$ = \pT/2 (dashed 
line), \pT (solid line), 2\pT (dotted line). The vertical bars show 
statistical uncertainties and square brackets are for systematic 
uncertainties. Not shown are 10\% absolute luminosity uncertainties. 
Panels (b) and (d) show comparisons of data and calculations.}
\label{fig:xsect}
\end{figure*}

In addition, we also measured the isolated direct-photon cross section 
with isolation criteria on the photon candidates, which can largely 
reduce the contributions from parton fragmentation and hadron decays. 
For any other particles within a cone of radius $r_{{\rm cone}} = 
\sqrt{(\delta\eta)^2 + (\delta\phi)^2} = 0.5$ of the signal photon, the 
sum of their energies is required to be less than 10\% of the energy of 
the signal photon: $E_{{\rm cone}} < 0.1 E_{\gamma}$. The energies of 
the neutral particles that pass charge-veto criteria were measured by 
the EMCal with a minimal threshold of 300 MeV. The momenta of the 
charged particles were measured by the drift chambers with a minimal 
threshold of 200 MeV/c. The efficiency of isolation criteria due to 
limited detector acceptance was corrected by using {\sc pythia}-simulated 
direct-photon events with the same isolation criteria as in the data. 
Similar to Eq.~(\ref{eq:inc}), the isolated direct-photon yield can be 
expressed as

\begin{equation} \label{eq:iso}
\gamma_{{\rm dir}}^{{\rm iso}} = \gamma_{{\rm total}}^{{\rm iso}} - \gamma_{\pi^0}^{{\rm iso}} - \left( R_{\pi^0}^{{\rm miss}} + V\delta_{h/\pi^0}^{\gamma} \right) \gamma_{\pi^0}^{{\rm isopair}},
\end{equation}

\noindent where $\gamma_{\pi^0}^{{\rm iso}}$ is the \pizero \ tagged-photon yield 
when each of the \pizero decay photons passes the isolation 
requirement. $\gamma_{\pi^0}^{{\rm isopair}}$ is the yield when a 
photon from a \pizero decay passes the isolation requirement while its 
partner photon energy is not included in the isolation-cone energy sum. 
Therefore, $R_{\pi^0}^{{\rm miss}}\gamma_{\pi^0}^{{\rm isopair}}$ 
represents the yield of \pizero decay photons that are missing the energy
of their partner photons. 
Similarly, the term $\delta_{h/\pi^0}^{\gamma}\gamma_{\pi^0}^{{\rm isopair}}$ 
corrects for the 
photons from other hadron decays that pass the isolation requirement 
while the energy of the partner photon is not included in the isolation cone 
energy sum.  To include the effect that one of the decay photons is 
vetoed by its partner decay photon due to isolation criteria, we use 
single $\eta$ and detector simulations to calculate the ratio of $\eta$ 
decay photons with and without isolation criteria, 
$V=\gamma^{{\rm iso}}_{\eta}/\gamma^{{\rm inc}}_{\eta}$, which varies 
from 0.01 to 0.1 depending on \pT.



The direct-photon cross section is calculated as

\begin{equation} \label{eq:xsecex}
E\frac{d^3\sigma}{dp^3} = \frac{1}{\mathcal{L}} \cdot \frac{1}{2\pi p_T} \cdot \frac{1}{\Delta p_T \Delta y} \cdot \frac{r_{{\rm pileup}} \cdot \gamma_{{\rm dir}}}{\epsilon},
\end{equation}

\noindent where $\epsilon$ includes corrections for the detector acceptance, 
photon reconstruction efficiency, trigger efficiency, and detector 
smearing effects and $r_{{\rm pileup}}$ is the correction for the pileup 
effects due to the large signal-integration time of the EMCal coupled 
with the high collision rate. It is approximately 0.8 (0.9) for 
inclusive (isolated) direct photons. The correction is obtained by a 
logarithmic extrapolation of the number of photons per event to zero 
event rate. The $\mathcal{L}$ is the integrated luminosity used for the 
analyzed data, and $\Delta y$ is the rapidity range.

The main systematic uncertainty sources are from the global energy scale 
of tuning the \pizero mass-peak position and energy nonlinearity of the 
EMCal response at high \pT. These are calculated by a single \pizero or 
photon generator with a fast detector simulation and depending on \pT 
were determined to be 14\%--19\% (7\%--13\%) for the inclusive (isolated) 
direct-photon cross section. The systematic uncertainties due to \pizero 
yield extraction and relative fractions of other hadron decays over 
\pizero are 2\%--12\% (0.5\%--2.5\%) and 5\%--14\% (0.4\%--6.0\%) for 
the inclusive (isolated) direct-photon cross section. These 
contributions for the isolated direct-photon cross section are 
relatively small compared to the inclusive case as the isolation 
requirement largely reduces these backgrounds. The loss of photons from 
conversions in the material before the EMCal is estimated using a 
single-photon generator plus full {\sc geant} detector 
simulation~\cite{Brun:1994aa}. The material of the vertex 
tracker~\cite{SONDHEIM2012993} leads to a (12.8 $\pm$ 1.9)\% probability 
for a photon to convert. This systematic uncertainty only contributes to 
the west arm, because in 2013 the east arm did not have a vertex-tracker 
installed.  Conversions in other materials lead to photon losses of 
$(3{\pm}1)$\% in the PbSc and $(4.5{\pm}1.3)$\% in the PbGl.  When 
calculating the direct-photon yield in Eq.~(\ref{eq:inc}) and 
Eq.~(\ref{eq:iso}), we vary the photon-conversion rate by its systematic 
uncertainty to get 1\%--8\% relative uncertainties of the direct-photon 
yield. The uncertainties from the EMCal detector resolution of 2\%--8\% 
and trigger of 2\%--4.5\% are also taken into account. Other 
uncertainties, including geometrical acceptance, trigger efficiencies, 
and pileup effect, are in total less than 7\%.

Figure~\ref{fig:xsect}(a) shows the measured inclusive direct-photon 
cross section at midrapidity in $\vec{p}+\vec{p}$ collisions at 
$\sqrt{s}=510$ GeV compared with NLO pQCD 
calculations~\cite{PhysRevD.48.3136,PhysRevD.50.1901} using NNPDF3.0 
parton-distribution functions (PDF)~\cite{Ball2015,Bonvini2015} and 
Gl{\"u}ck-Reya-Vogt (GRV) fragmentation functions 
(FF)~\cite{PhysRevD.45.3986}. The pseudorapidity range for this 
measurement is $|\eta|<0.25$ after the fiducial requirement that removes 
edge towers of the EMCal. The calculation is in good agreement with the 
data within the uncertainties for $p_T>12$~GeV/$c$, but underestimates 
the yield by up to a factor of $\approx$3 for \mbox{$p_T<12$ GeV/$c$}. 
This discrepancy is possibly due to multiparton interactions and parton 
showers~\cite{Nason_2004,Frixione_2007,Alioli2010,Jezo2016,Klasen2018}. 
The isolated direct-photon cross section is shown in 
Fig.~\ref{fig:xsect}(c) as a function of \pT and compared with the NLO 
pQCD calculation~\cite{PhysRevD.48.3136,PhysRevD.50.1901} using 
NNPDF3.0~\cite{Ball2015,Bonvini2015} and 
GRV~FF~\cite{PhysRevD.45.3986}. The calculation is in good agreement with 
the data within the uncertainties, with slight overestimation in the 
lowest \pT bins.


The double-helicity asymmetry is defined as

\begin{equation} \label{eq:all}
A_{LL} = \frac{\Delta\sigma}{\sigma} = \frac{\sigma_{++}-\sigma_{+-}}{\sigma_{++}+\sigma_{+-}},
\end{equation}

\noindent where $\sigma_{++}$ ($\sigma_{+-}$) is the cross section for the same 
(opposite) helicity proton-proton collisions. This can be rewritten in 
terms of particle yield and beam polarizations:

\begin{equation}
A_{LL} = \frac{1}{P_BP_Y} \frac{N_{++}-RN_{+-}}{N_{++}+RN_{+-}}
\end{equation}

\noindent where $N_{++}$ ($N_{+-}$) is the number of isolated direct photons from 
the collisions with the same (opposite) helicities. $P_{B}$ ($P_{Y}$) 
are the polarizations for the blue (yellow) proton beams, and the 
average values in 2013 were 0.55 (0.57)~\cite{POBLAGUEV2020164261}. $R 
=(\mathcal{L_{++}}/\mathcal{L_{+-}})$ is the relative luminosity that is 
measured by the BBC. The systematic contribution of $R$ to $A_{LL}$ was 
found to be $3.8{\times}10^{-4}$~\cite{PhysRevD.93.011501}.

The asymmetry was calculated for photon candidates that passed the same 
time-of-flight, minimum-energy, and isolation requirements as in the 
cross-section analysis. A z-vertex requirement of 30 cm is used for 
the asymmetry measurement. The asymmetry contribution for background 
photons from \pizero's decay was calculated from the sideband regions 
(47--97 MeV/c$^2$ and 177--227 MeV/c$^2$) below and above the \pizero 
mass peak (112--162 MeV/c$^2$) using the inclusive photon sample due to 
the limited statistics in the isolated photon sample. The asymmetry for 
other hadron decays (mostly $\eta$ decays) was taken as $A_{LL}^{\eta}$ 
from previous PHENIX measurements at 
$\sqrt{s}=200$~GeV~\cite{PhysRevD.90.012007} by assuming $x_T$ scaling. 
The difference in $A^{\eta}_{LL}$ between 200 GeV and 510 GeV for a 
given $x_{_{T}}$ is expected to be much smaller than the experimental 
uncertainty of the 200 GeV result which was used to assign a systematic 
uncertainty~\cite{PhysRevLett.113.012001, 2014276}. The 
background-corrected asymmetry can be calculated as

\begin{equation} \label{eq:all-dir}
A_{LL}^{{\rm dir}} = \frac{A_{LL}^{{\rm total}} - r_{\pi^0}A_{LL}^{\pi^0} -r_h A_{LL}^{\eta}}{1 - r_{\pi^0} - r_h},
\end{equation}

\noindent where $r_{\pi^0}$ (10\%--14\%) and $r_h$ (0.6\%--1.4\%) are background 
fractions of \pizero and other hadron-decay photons, respectively.
We used a bunch-shuffling 
technique which assigned a random spin polarization to 
each bunch and examined the distribution of resulting asymmetries
ensure there were no false asymmetries arising from 
unknown systematic effects~\cite{PhysRevD.90.012007}.
The data were divided into subgroups according to the 
bunch spin patterns that were used to fill the RHIC rings, and 
calculated asymmetries were found to be consistent.

\begin{figure}[htb]
\includegraphics[width=1.0\linewidth]{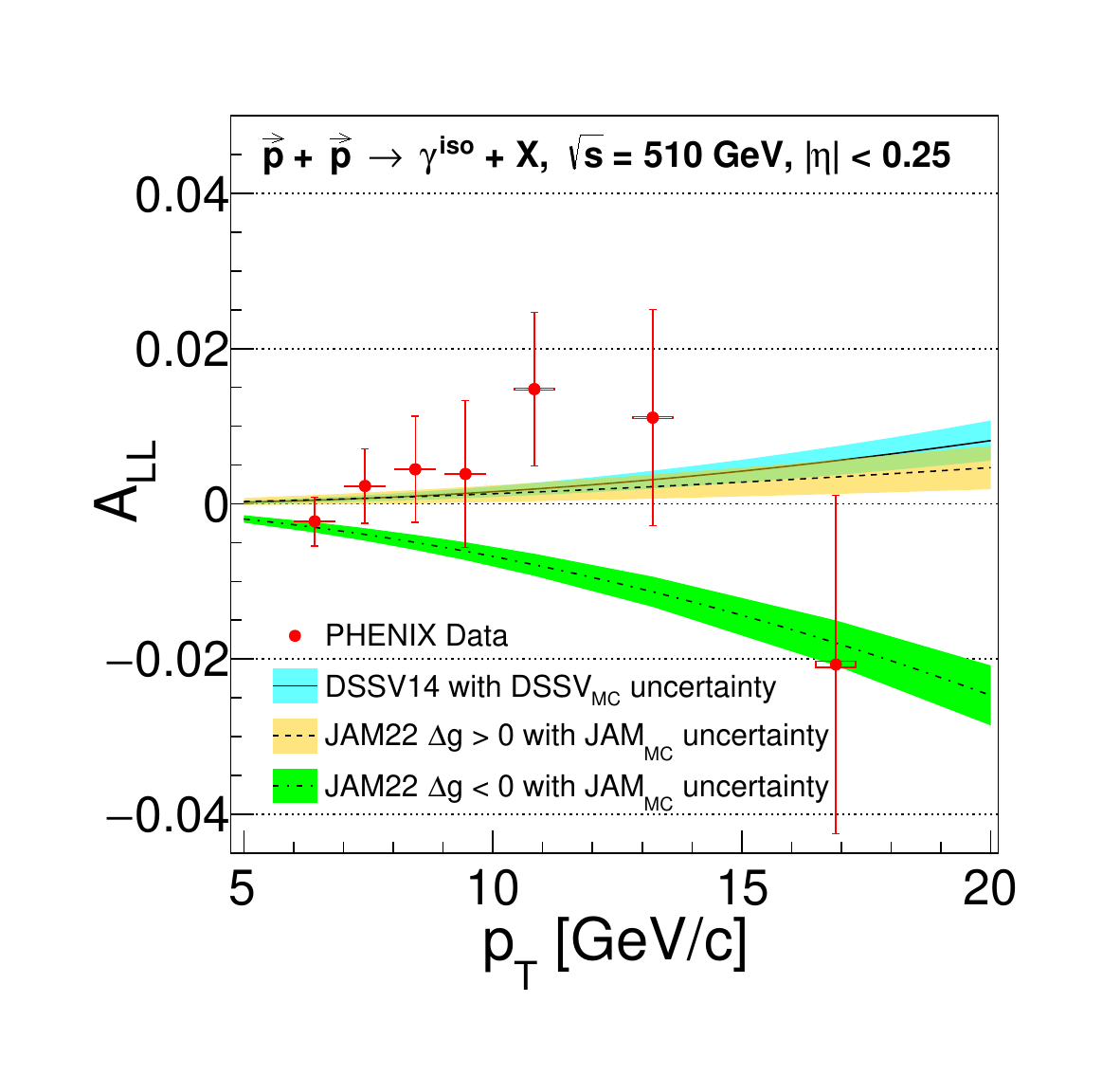}
\caption{Double-helicity asymmetry $A_{LL}$ $vs$ $p_{T}$ for isolated 
direct-photon production in polarized $p$$+$$p$ collisions at 
$\sqrt{s}=510$ GeV at midrapidity. Vertical error bars (boxes) represent 
the statistical (systematic) uncertainties. The systematic uncertainties
for $p_T < 10$ GeV/c are smaller than the marker size. Not shown are
a $3.9{\times}10^{-4}$ shift uncertainty from relative luminosity and 
a 6.6\% scale uncertainty from polarization. The DSSV14 and JAM22 
calculations are shown with $1\sigma$ uncertainty bands obtained from MC 
replicas~\protect\cite{PhysRevLett.101.072001,PhysRevLett.113.012001,PhysRevD.100.114027,Zhou:2022wzm,PhysRevD.106.L031502}.
JAM22 calculations are based on PDF sets from the global analysis of 
the JAM Collaboration~\cite{PhysRevD.106.L031502}, and the code to 
calculate the asymmetries was provided by W. Vogelsang.
}
\label{fig:all}
\end{figure}

Figure~\ref{fig:all} shows the double-helicity asymmetry of isolated 
direct-photon production in longitudinally polarized proton-proton 
collisions at $\sqrt{s}=510$~GeV for $6<p_T<20$~GeV/$c$.
The corresponding gluon momentum fraction is $x \approx 2p_T/\sqrt{s}$.
In the asymmetry  measurement, systematic effects are largely canceled.
The systematic uncertainties in Fig.~\ref{fig:all} include point-to-point
uncertainties from background estimation and false asymmetries in the
background due to pileup effects at low \pT.
The NLO pQCD calculation was obtained using the DSSV14 polarized PDFs,
the NNPDF3.0 unpolarized PDFs and the GRV~FF for the renormalization and factorization 
scales $\mu=p_T$ with the $1\sigma$ uncertainty band determined via MC replicas (a 
sampling variant of the DSSV14 set of helicity parton 
densities)~\cite{PhysRevLett.101.072001,PhysRevLett.113.012001, 
PhysRevD.100.114027}. The calculation is in good agreement with the 
results, within experimental uncertainties.

The two dashed curves in Fig.~\ref{fig:all} come from the global 
analysis of the JAM 
Collaboration~\cite{Zhou:2022wzm,PhysRevD.106.L031502}. They 
found there are two distinct sets of solutions for the polarized 
gluon PDF, $\Delta g$, which differ in sign. Even though the 
solutions with $\Delta g<0$ violate the positivity assumption, 
$|\Delta g|<g$, all previous data cannot exclude those solutions due 
to the mixed contributions from quark-gluon and gluon-gluon 
interactions. However, the direct-photon \ALL comes mainly from the 
quark-gluon interactions and has $\chi^2 =$ 4.7 and 12.6 for 7 data 
points for the $\Delta g>0$ and $\Delta g<0$ solutions, respectively, 
with the difference of 7.9 between $\chi^2$ values implying that the 
negative solution is disfavored at more than 2.8$\sigma$ level.

In summary, PHENIX has measured the cross section and \ALL of direct 
photons at midrapidity in $\vec{p}+\vec{p}$ collisions at 
$\sqrt{s}=510$~GeV. The NLO pQCD calculations are consistent with the 
results except at lower \pT where the calculations underestimate the 
inclusive direct-photon cross section. With isolation criteria, the 
partonic level calculation is in better agreement with the measurement. 
This is the first measurement of the \ALL of direct photons, which is 
sensitive to the polarized-gluon distribution inside the proton.
Our data are well consistent with the positive gluon-spin contributions
and strongly disfavor the negative gluon-spin scenario, that the
previously published data were unable to resolve.



We thank the staff of the Collider-Accelerator and Physics
Departments at Brookhaven National Laboratory and the staff of
the other PHENIX participating institutions for their vital
contributions. We also thank W. Vogelsang for providing
the code for direct photon asymmetry calculations.
We acknowledge support from the Office of Nuclear Physics in the
Office of Science of the Department of Energy,
the National Science Foundation,
Abilene Christian University Research Council,
Research Foundation of SUNY, and
Dean of the College of Arts and Sciences, Vanderbilt University
(U.S.A),
Ministry of Education, Culture, Sports, Science, and Technology
and the Japan Society for the Promotion of Science (Japan),
Natural Science Foundation of China (People's Republic of China),
Croatian Science Foundation and
Ministry of Science and Education (Croatia),
Ministry of Education, Youth and Sports (Czech Republic),
Centre National de la Recherche Scientifique, Commissariat
{\`a} l'{\'E}nergie Atomique, and Institut National de Physique
Nucl{\'e}aire et de Physique des Particules (France),
J. Bolyai Research Scholarship, EFOP, the New National Excellence
Program ({\'U}NKP), NKFIH, and OTKA (Hungary),
Department of Atomic Energy and Department of Science and Technology
(India),
Israel Science Foundation (Israel),
Basic Science Research and SRC(CENuM) Programs through NRF
funded by the Ministry of Education and the Ministry of
Science and ICT (Korea),
Physics Department, Lahore University of Management Sciences (Pakistan),
Ministry of Education and Science, Russian Academy of Sciences,
Federal Agency of Atomic Energy (Russia),
VR and Wallenberg Foundation (Sweden),
University of Zambia, the Government of the Republic of Zambia (Zambia),
the U.S. Civilian Research and Development Foundation for the
Independent States of the Former Soviet Union,
the Hungarian American Enterprise Scholarship Fund,
the US-Hungarian Fulbright Foundation,
and the US-Israel Binational Science Foundation.



%
 
\end{document}